\documentclass[prl,amssymb,twocolumn,showpacs,superscriptaddress,floatfix,letterpaper]{revtex4}

\usepackage{graphicx}
\usepackage{amsmath}

\begin{document}

\title{Oscillations in two-species models: tying the stochastic and deterministic approaches}

\author{Sebasti\'{a}n Risau-Gusman}
\affiliation{Centro At\'{o}mico Bariloche and CONICET, 8400 S. C. de Bariloche, Argentina}

\author{Guillermo Abramson}
\affiliation{Centro At\'{o}mico Bariloche, CONICET and Instituto
Balseiro, 8400 S. C. de Bariloche, Argentina}

\date{\today}

\begin{abstract}
We analyze general two-species stochastic models, of the kind
generally used for the study of population dynamics. We show that
the conditions for the stochastic (microscopic) model to display
approximate sustained oscillatory behavior are governed by the
parameters of the corresponding deterministic (macroscopic) model.
We provide a quantitative criterion for the quality of the
stochastic oscillation, using a dimensionless parameter that depends
only on the deterministic model. When this parameter is small, the
oscillations are clear, and the frequencies of the stochastic and
deterministic oscillations are close, for all stochastic models
compatible with the same deterministic one. On the other hand, when
it is large, the oscillations cannot be distinguished from a noise.
\end{abstract}

 \maketitle

\enlargethispage*{\baselineskip}

\begin{figure}[b]
\begin{center}
\resizebox{\columnwidth}{!}{\includegraphics{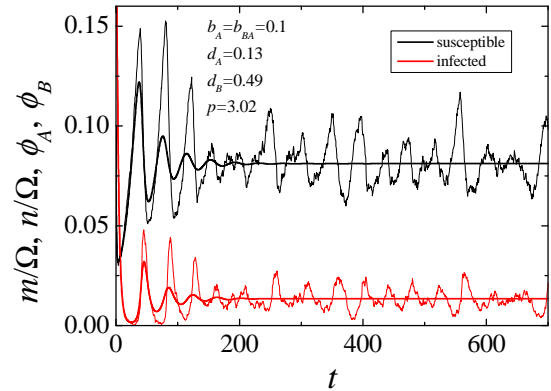}}
\caption{Deterministic dynamics (smooth lines) and one stochastic
realization (fluctuating lines) of an SI epidemic model (susceptible
and infected, respectively A and B). The dynamics includes birth and
death processes in both populations, and contagion. A self-limiting
intraspecific competition mechanism is implemented as
in~\cite{mckane04,mckane05}, with a total system size
$\Omega=5\!\times\! 10^5$. Transition rates are:
$T_{10}=2(b_A\phi_A+b_{B\!A}\,\phi_B)(1-\phi_A-\phi_B)$,
$T_{-10}=d_A\phi_A$, $T_{0-1}=d_B\phi_B$, $T_{-11}=2p\phi_A\phi_B$
(see Eq.~(\ref{rates})).}
\label{evolucion}
\end{center}
\end{figure}

It is well known that the dynamics of many systems of two species
can display an oscillatory behavior in the populations of both
agents. This happens in predator-prey systems~\cite{begon}, in
models of measles epidemics~\cite{wilson45}, in chemical systems
such as those exemplified by the Brusselator~\cite{prigogine68},
etc. These systems are usually modelled by a set of two coupled
ordinary differential equations, which are assumed to represent a
macroscopic level of description of the system. Oscillations can
appear in these models as limit-cycle solutions to the equations.
However, it frequently happens that the macroscopic model only has
\emph{damped} oscillatory solutions, even though the modelled system
displays sustained oscillations in its populations in the same
region of parameter values. Examples of this are not uncommon in
population dynamics (see, e.g., the discussion in~\cite{renshaw}
with regard to predator-prey and measles problems). It has often
been noted that the stochastic counterpart of these models---assumed
to represent a more microscopic level description of the same
system--- usually do display a kind of \emph{sustained} oscillatory
behavior, with a frequency very similar to the one of the damped
solutions of the differential
equations~\cite{bartlett57,hethcote89,solari01} (see
Fig.~\ref{evolucion} for an example based on a susceptible-infected
epidemic model). These oscillations are said to be generated by the
demographic, or intrinsic, noise~\cite{mckane05}. The problem is
that stochasticity precludes a clear-cut definition of
``oscillations" for such systems. Therefore, the comparison between
the results of the stochastic and deterministic approaches is often
made on a qualitative basis.

In this Letter we address the problem of sustained oscillations in
stochastic models, in an attempt to characterize the oscillatory
regime. We show that the conditions for well defined oscillations
are given by the parameters of the corresponding macroscopic
(deterministic) model only, disregarding the details of the
microscopic (stochastic) one. In other words, this general result
proves that the conditions are the same for \emph{any} stochastic
model that corresponds to the same macroscopic one. To this end, we
define a criterion of ``quality" of oscillatory behavior for a large
class of stochastic two-species systems and we show, by means of a
van Kampen expansion of the master equation, that the information
given by the deterministic system---embodied in a deterministic
parameter---is enough to provide good bounds on this quality. In
other words, we show that the quality of the oscillations is only
weakly dependent on the details of the demographic noise. Moreover,
it is shown that oscillations become clear if and only if the
deterministic parameter vanishes. We also suggest a heuristic value
for the quality below which one can be almost certain that the
evolution of both populations does ``look" oscillatory.

We consider systems of two populations, A and B, described by
stochastic variables $m(t)$ and $n(t)$. The state of the system is
defined by the joint probability $P(m,n;t)$ that the system has $m$
individuals of species A, and $n$ individuals of species B. The
transition from a state with $(m,n)$ individuals to a state with
$(m+i,n+j)$ individuals takes place at a rate:
\begin{equation}
T(m+i,n+j|m,n)=f(\Omega)T_{ij}(\frac{m}{\Omega},\frac{n}{\Omega}),
\label{rates}
\end{equation}
where $-k<i<k$ and $-k<j<k$. $\Omega$ is a scale parameter that
governs the fluctuations of the stochastic evolution. Its precise
definition depends on the system, but one chooses it in such a way
that for large $\Omega$ the fluctuations are small. It usually
represents the volume containing the reactants in chemical
systems~\cite{vankampen}, or the available resources in biological
ones~\cite{mckane05}. The constant $k$ gives the maximal number of
elements that can appear, or disappear, from a given population at
each step of the dynamics. The most common choice are one-step
processes, with $k=1$.

\enlargethispage*{\baselineskip}

The evolution of the probability $P(m,n;t)$ is given by the master
equation~\cite{vankampen}:
\begin{eqnarray}
\frac{\partial P(m,n;t)}{\partial t} &=& \sum_{ij} P(m-i,n-j;t) \,
T_{ij}(\frac{m-i}{\Omega},\frac{n-j}{\Omega}) \nonumber \\
&& - P(m,n;t) \sum_{ij} T_{ij}(\frac{m}{\Omega},\frac{n}{\Omega}),
\label{master}
\end{eqnarray}
where, as in the rest of this Letter, the summation indices run from
$-k$ to $k$.

Except for a few simple cases, this equation is extremely difficult
to solve exactly. For this reason many methods have been devised to
look for approximate solutions. Perhaps the best known, and most
applied, is the van Kampen expansion~\cite{vankampen}. In the
following we sketch the main steps leading to the series solution (a
detailed account can be found in van Kampen's
book~\cite{vankampen}).

If one assumes that, at time zero, the system is in a state where
both populations have well defined macroscopical values,
$P(m,n)=\delta(m-m_0)\delta(n-n_0)$, with the initial values of
order $O(\Omega)$, it is reasonable to expect that at later times
$P(m,n)$ will have a sharp peak at some position of order
$O(\Omega)$ (in both populations), and a width of order
$O(\Omega^{1/2})$. That is, the fluctuating populations will satisfy
$m=\Omega \phi_A+\sqrt{\Omega} \xi_A$ and $n=\Omega
\phi_B+\sqrt{\Omega} \xi_B$, where the variables $\phi$ represent
the ``macroscopic'' evolution, while the stochastic variables $\xi$
represent fluctuations around them. Replacing this in
Eq.~(\ref{master}), equating terms of the same order in $\Omega$ and
adequately rescaling the time, one obtains, for the leading order:
\begin{eqnarray}
\dot{\phi}_A &=& \sum_{ij} i\, T_{ij}(\phi_A,\phi_B) \equiv C_A (\phi_A,\phi_B), \nonumber \\
\dot{\phi}_B &=& \sum_{ij} j\, T_{ij}(\phi_A,\phi_B) \equiv C_B
(\phi_A,\phi_B).
\label{first}
\end{eqnarray}
These equations, called \emph{deterministic} or \emph{macroscopic},
are usually the starting point of many models of chemical and
biological systems. They are generally written down from macroscopic
considerations of the population dynamics, disregarding its
individual level origin. To analyze the differences between the
stochastic (individual level) and the deterministic (population
level) approaches one usually chooses a stochastic model that gives
the right deterministic equations. In the limit of infinite size
($\Omega\to\infty$), Eqs.~(\ref{master}) are also satisfied by the
average  populations.

The deterministic equilibria are obtained by solving the system
$C_A(\phi_A,\phi_B)=C_B(\phi_A,\phi_B)=0$, and their stability is
studied by means of a linear stability analysis. When the system is
close to a stable equilibrium, its evolution can be approximated by
that of a damped oscillator. In the underdamped regime, the damping
factor and the frequency of oscillation are, respectively:
\begin{eqnarray}
\gamma &=& \Delta \epsilon/2, \nonumber \\
\omega_d^2 &=& \Delta (1- \epsilon^2/4),
\label{deterministic}
\end{eqnarray}
with
\begin{equation}
\epsilon = |T|/ \sqrt{\Delta},
\end{equation}
where $\Delta$ is the Jacobian of $\vec{C}$, and $T$ its trace:
\begin{eqnarray}
\Delta &=& C_{A,A} C_{B,B} - C_{A,B} C_{B,A} , \nonumber \\
T &=& C_{A,A} + C_{B,B},
\label{deltat}
\end{eqnarray}
where $C_{i,j}=\frac{\partial C_i}{\partial \phi_j}$. The
underdamped regime is therefore given by the condition $\epsilon<2$.
We show below that this parameter, which depends only on the
parameters of the macroscopic Eq.~(\ref{first}), plays a fundamental
role in the characterization of the oscillations of stochastic
origin. Notice also that the number of oscillations observed in the
characteristic time $\gamma^{-1}$ depends only on $\epsilon$ (for
small $\epsilon$, it is just $2/\epsilon$).

The following order in the van Kampen expansion gives the evolution
of $\Pi(\xi_A,\xi_B,t)$, the joint probability function of the
fluctuations, in the form of a Fokker-Planck equation. To look for
oscillations in the fluctuations it is easier to work with the
equivalent Langevin equations, as shown by McKane~\cite{mckane05}:
\begin{eqnarray}
\dot{\xi}_A &=& -C_{A,A} \xi_A -C_{A,B} \xi_B + L_A (t) \nonumber \\
\dot{\xi}_B &=& -C_{B,A} \xi_A -C_{B,B} \xi_B + L_B (t)
\label{langevin}
\end{eqnarray}
where $L_A(t)$ and $L_B(t)$ are delta-correlated Gaussian noises of
zero mean, satisfying $\langle L_A(t) L_A(t') \rangle =
D_A\delta(t-t')$, $\langle L_B(t) L_B(t') \rangle =
D_B\delta(t-t')$, and $\langle L_A(t) L_B(t') \rangle =
D_{AB}\delta(t-t')$. The noise intensities are given by:
\begin{eqnarray}
D_A(\phi_A,\phi_B)&=&\sum_{i,j} i^2\, T_{ij} (\phi_A,\phi_B), \nonumber \\
D_B(\phi_A,\phi_B)&=&\sum_{i,j} j^2\, T_{ij} (\phi_A,\phi_B), \\
D_{AB}(\phi_A,\phi_B)&=&\sum_{i,j} ij\, T_{ij} (\phi_A,\phi_B).
\nonumber
\label{gs}
\end{eqnarray}

By Fourier transforming Eqs.~(\ref{langevin}) it is straightforward
to obtain the power spectrum of the fluctuations around the
deterministic equilibrium~\cite{mckane05}. In the following we
concentrate on population A. The corresponding expressions for
population B are obtained by exchanging $A$ and $B$ in all the
subindices. The average power spectrum of $\xi_A$ is
\begin{equation}
\langle S_A(\omega) \rangle =
\frac{F_A+\hat{\omega}^2}{(1-\hat{\omega}^2)^2+\hat{\omega}^2
\epsilon^2},
\label{powspec}
\end{equation}
where
\begin{eqnarray}
\hat{\omega}^2 \!\!&=&\!\! \omega^2/\Delta, \nonumber \\
F_A \!\!&=&\!\! \frac{C_{A,B}^2 D_{B}+C_{B,B}^2 D_{A}-2 C_{A,B}
C_{B,B} D_{AB}}{\Delta D_{A}}.
 \label{defs}
\end{eqnarray}
We stress that $F_A$ (and correspondingly $F_B$), through its
dependence on $C_A$, $C_B$, $D_A$ and $D_B$, depends ultimately on
the transition probabilities that define the model.
\begin{figure}[t]
\begin{center}
\resizebox{\columnwidth}{!}{\includegraphics{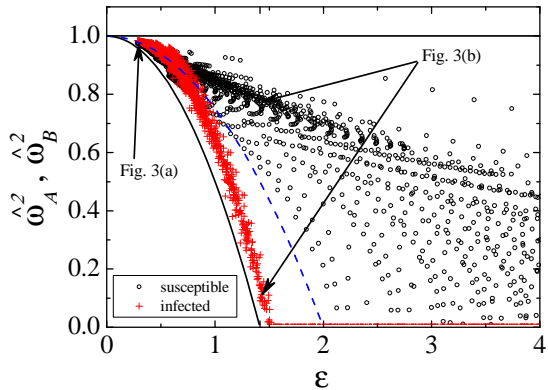}}
\caption{$\hat{\omega}_A^2$ and $\hat{\omega}_B^2$ for the same SI
model as in Fig.~\ref{evolucion}. The points correspond to a uniform
scanning of a portion of phase space: $b_A=b_{BA}=0.1$, $d_A\in
(0,0.2)$, $d_B\in (0,0.5)$, $p\in(2d_B,3.2)$. The full lines show
the bounds of Eq.~(\ref{boundw}), while the dashed one corresponds
to $\omega_d$. The arrows point to the values corresponding to the
parameters used in Fig.~\ref{timeseries}.}
\label{omega}
\end{center}
\end{figure}

\enlargethispage*{\baselineskip}

It is straightforward to see that $\langle S_A(\omega) \rangle$ is
either monotonically decreasing or it has a single maximum at
\begin{equation}
\hat{\omega}_A^2 = -F_A+ \sqrt{(F_A+1)^2 - \epsilon^2 F_A}.
\label{omegaA}
\end{equation}
The condition of positivity for the argument of the square root
gives the region in phase space where the power spectrum has a
single maximum. Notice that for $\epsilon < \sqrt{2}$ this condition
is fulfilled \emph{regardless of the exact dependence of $F_A$ on
the parameters of the model}. It can also be proved that
$\hat{\omega}_A^2$ satisfies~\cite{risau07}:
\begin{equation}
1-\epsilon^2/2<\hat{\omega}_A^2<1
\label{boundw}
\end{equation}
(these bounds seem to be tight). In particular, this implies that in
all the possible stochastic models that lead to the same
deterministic equations (same $C$'s, different $D$'s) the position
of the maximum can only vary within a finite range, that shrinks
with $\epsilon$.

For small $\epsilon$, $\hat{\omega}_A^2$ tends to $1$, which means
that not only the frequencies of possible oscillations for both
populations become close, but also that they become close to
$\omega_d$, the frequency of the damped oscillations of the
deterministic model (which also tends to 1 as $\epsilon\to 0$). It
is in this regime that the populations show the coherent dynamics
characteristic of stochastic oscillations. This motion will be
further characterized below by the quality of the spectrum peak.
Figure \ref{omega} shows $\hat{\omega}^2_A$ and $\hat{\omega}^2_B$
as functions of $\epsilon$ for the SI model presented in
Fig.~\ref{evolucion}, for a wide range of system parameters. The
bounds given by Eq.~(\ref{boundw}) are shown by continuous lines.
Each point represents the normalized squared frequency for one set
of parameters, for both populations. The deterministic frequency,
$\omega_d$, is also shown, to emphasize the difference between the
three frequencies present in the system.

When $\sqrt{2}< \epsilon <2$ there can be some stochastic models for
which no peak is present in $S(\omega_A)$ or $S(\omega_B)$. And, for
some values of $F_A$ or $F_B$, it can happen that the power spectrum
of any population has a maximum even if $\epsilon>2$, i.e.~even when
the deterministic system does not display damped oscillations (see
Fig.~\ref{omega}: all the points to the right of $\epsilon\!=\!2$
correspond to systems with a peak in the spectrum of the susceptible
(A) population, no peak in the infected (B) one, and no damped
oscillations in the deterministic model). These two features show
that the peaks of the stochastic power spectrum on the one hand, and
the deterministic damped oscillations on the other, are not
necessarily closely related.

The above discussion establishes the conditions for the existence of
a peak in the power spectrum of one or both populations. That is,
for the existence of a preferred frequency in their dynamics. But,
should all peaks in the power spectrum be regarded as
``oscillations''? The answer to this question is certainly negative,
and leads one to look for a criterion to quantify how close a time
series is to an oscillatory movement. This can be done by defining
the ``quality'' of the oscillation as a measure of the sharpness of
the peak. We propose one such measure in the following.

Given a power spectrum of the form~(\ref{powspec}) we define the
quality of a peak at $\omega_{peak}$ as
\begin{equation}
Q_A(\omega)= \frac{\omega_{peak} \langle S_A(\omega_{peak}) \rangle
}{ \int \langle S_A(\omega) \rangle d \omega }.
\label{defQA}
\end{equation}
This quantity is dimensionless and scale invariant. It is related to
Fisher's kappa, which measures the non-stationarity of a signal,
given its periodogram~\cite{fisher29}. For functions with only one
peak, $Q_A$ increases as the peaks grows. For power spectra of the
form~(\ref{powspec}), $Q_A$ can be readily calculated (using that
$\int \langle S_A(\omega) \rangle d \omega = \langle \xi_A^2
\rangle$, see~\cite{vankampen}):
\begin{equation}
Q_A(\omega_A)=
\frac{\hat{\omega}_A \epsilon}
     {(\hat{\omega}_A^2-1)^2+\hat{\omega}_A^2\epsilon^2}
\left(
\frac{F_A+\hat{\omega}_A^2}{F_A+1}
\right) .
\label{QA}
\end{equation}

\begin{figure}[t]
\begin{center}
\resizebox{\columnwidth}{!}{\includegraphics{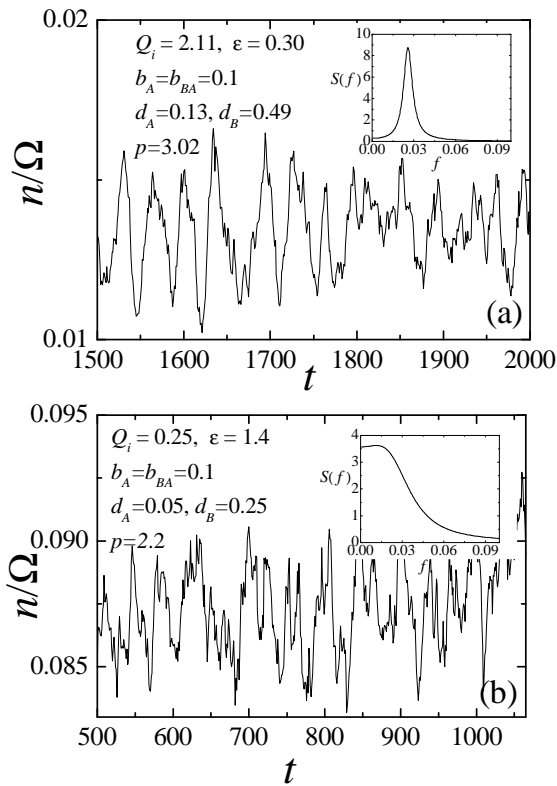}} \caption{Two
stochastic realizations of the SI model, with different qualities.
The insets show the corresponding analytical average power spectra.
Only the infected population is shown. The arrows in
Fig.~\ref{omega} point to the corresponding frequencies:
$\omega_A\sim\omega_B\sim\omega_d$ for the good quality case shown
in (a), and $\omega_A\gg\omega_B$ for the bad quality one shown in
(b). $\Omega=10^5$.}
\label{timeseries}
\end{center}
\end{figure}

The quality $Q_A$ diverges as $\epsilon$ vanishes, regardless of
the exact dependence of $F_A$ on the parameters of the model.
Therefore, one can assure that the corresponding time series will
look oscillatory when $\epsilon$ is sufficiently small (see
Fig.~\ref{timeseries} for an example of this). In such a case, we
have already shown that the frequencies of both populations are very
close, and also very close to the frequency of the deterministic
damped oscillations.

\begin{figure}[t]
\begin{center}
\resizebox{\columnwidth}{!}{\includegraphics{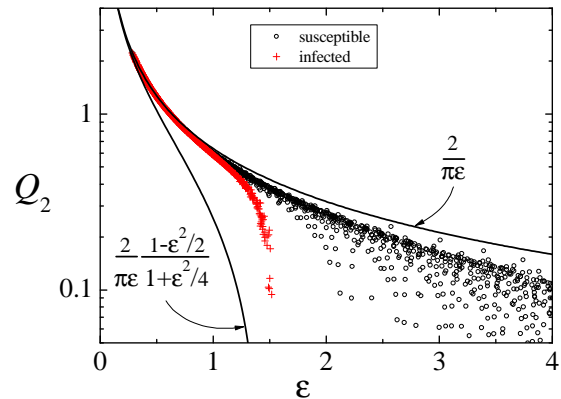}}
\caption{$Q_A$ and $Q_B$ as functions of $\epsilon$ for the SI model
of Fig.~\ref{evolucion}. The points correspond to the same portion
of phase space as in Fig.~\ref{omega}. The lines show the upper and
lower bounds of Eq.~(\ref{boundsQA}).}
\label{figuQ}
\end{center}
\end{figure}

Could it also happen that, for large values of $\epsilon$, when the
frequencies of populations $A$ and $B$ can be rather different, one
gets very sharp peaks? It can be shown that this is not the case by
giving bounds of $Q_A$ that depend solely on
$\epsilon$~\cite{risau07}:
\begin{equation}
\frac{2}{\pi\epsilon}
\left(
\frac{1-\epsilon^2/2}{1+\epsilon^2/4}
\right)
< Q_A(\omega) <
\frac{2}{\pi\epsilon}.
\label{boundsQA}
\end{equation}
The upper bound shows that, when $\epsilon$ is not small, the peak
cannot be arbitrarily sharp. On the other hand, the lower bound
shows that, when $\epsilon$ is small the peak is sharp for {\it all}
the stochastic counterparts of a deterministic model. In
Fig.~\ref{figuQ} we illustrate this by showing several values of
$Q_A$ and $Q_B$ for the SI model, along with the corresponding
bounds.

\enlargethispage*{\baselineskip}

One practical question remains: what is the critical quality value
above which one can be sure that the time series will indeed
``look'' oscillatory? As it is to be expected, the continuous nature
of $Q$ precludes a conclusive answer. From exhaustive observations
of different models we find that when $Q(\omega_{peak}) >1$, the
oscillations are well defined and notably different from a noisy
evolution (see Fig.~\ref{figuQ}).

In summary, we have shown that, by defining a quality measure, one
can quantify the ``oscillatory look'' of a time series.
Interestingly, we find that oscillations are present only when
$\epsilon$ is small. This means that, given a deterministic model,
one can know, using the bounds~(\ref{boundsQA}), whether the time
series given by {\it any} stochastic counterpart of the model will
look oscillatory or not. In addition, we have shown that, when
oscillations are clear, the corresponding frequencies of both
populations will be close to each other and to the frequency of
the damped oscillation of the deterministic system.

Given that our conclusions are based on the analysis of the first
two terms of the systematic van Kampen's expansion of the master
equation, they are exact only in the limit $\Omega\to \infty$. These
analytical results, nevertheless, compare well with the numerical
observations made on finite systems. More details about the validity
of the expansion for finite systems will be given
elsewhere~\cite{risau07}.

\begin{acknowledgments}
We are grateful to E. Andr\'{e}s, I. Peixoto, A. Aguirre, L. O\~{n}a and H.
Solari for valuable discussions. We acknowledge financial support
from ANPCyT (PICT-R 2002-87/2), CONICET (PIP 5414) and UNCUYO
(06/C209).
\end{acknowledgments}

\end{document}